\documentclass[12pt]{article}
\usepackage{graphicx, hyperref}

\begin{document}
	
%

\begin{center}
    \sf {\Large {\bfseries Effect of the dose distribution and organ architecture on the toxicity in FLASH radiotherapy: a modeling study}} \\
    \vspace*{10mm}
    Juan Pardo-Montero$^{1,2,*}$\\
    $^1$Group of Medical Physics and Biomathematics, Instituto de Investigación Sanitaria de Santiago (IDIS), Santiago de Compostela, Spain \\
    $^2$Department of Medical Physics, Complexo Hospitalario Universitario de Santiago, Santiago de Compostela, Spain
    \vspace{5mm}\\
    Version typeset \today\\
\end{center}

\pagenumbering{roman}
\setcounter{page}{1}
\pagestyle{plain}

* E-mail:  juan.pardo.montero@sergas.es \\

\vspace{10pt}

	
\begin{abstract}

\noindent\textbf{Objective}: This study aims to investigate the influence of organ architecture (specifically the distinction between serial and parallel tissue) on the protective FLASH effect when organs are irradiated with inhomogeneous dose distributions.

\noindent\textbf{Approach}: An \emph{in silico} modeling framework was developed using two distinct methods to calculate the effective FLASH dose: the first method utilized a biophysical model of radiolytic oxygen depletion (ROD); the second employed a phenomenological logistic function where the effective FLASH dose is a function of local dose and dose rate. Both models assume that the underlying mechanism behind the FLASH effect is local. Normal Tissue Complication Probability (NTCP) for heterogeneous dose distributions was calculated using the Lyman-Kutcher-Burman (LKB) model and the generalized equivalent uniform dose, varying the volume effect parameter $n$ from 1.0 (parallel) to below $0.01$ (serial) to explore different architectures. 

\noindent\textbf{Main results}: Both the ROD and phenomenological models showed FLASH sparing compared to conventional radiotherapy. Also, the sparing increased with decreasing $n$ (the sparing is more important for serial organs). For example, for a specific calculation, when the NTCP for conventional radiotherapy was 0.2 (set value) the corresponding NTCP for FLASH delivery ranged from 0.14 for $n$=1 to 0.11 for $n$=0.1.

\noindent\textbf{Significance}: Our results indicate that if the underlying mechanism/s behind the FLASH effect is/are local, the toxicity sparing associated to FLASH-RT can be dependent on the architecture of the irradiated organ/tissue, being more important for serial organs, which are more sensitive to large local doses than to average doses.
\end{abstract}

%
\vspace{2pc}

\section{Introduction} \label{section_intro}

FLASH radiotherapy (FLASH-RT) can spare normal tissue compared to conventional radiotherapy (CONV-RT), as shown by many \textit{in vivo}~\cite{favaudon2014, montay2017, montay2018, vozenin2019, montay2019, levy2020, diffenderfer2020, liljedahl2022, gao2022}. The mechanisms behind this sparing are still not entirely clear. Several studies suggest that the protective effect arises from the radiolytic oxygen depletion (ROD) caused by the ultra-high dose rate of FLASH-RT~\cite{favaudon2022, pratx2019a, pratx2019b, petersson2020, cao2021, el2022, ha2022}. Other studies suggest that oxygen depletion alone is insufficient to explain the sparing effect~\cite{boscolo2021, jansen2022, limoli2023}. Additional physicochemical factors have been investigated as the mechanism behing the FLASH effect, such as variations in reactive species production, the recombination of free radicals, or lipid metabolism ~\cite{spitz2019, jin2020, abolfath2023, ni2025, geirnaert2025}. Radiation-induced immune effects, in particular the sparing of circulating immune cells, has also been suggested as a possible mechanism of the FLASH effect~\cite{kim2024}.

Many of the experimental studies showing the FLASH effect deliver quite homogeneous doses of radiation to the organs/tissues under investigation. This is far from the clinical scenario, where the dose delivered to the tumor may be quite homogeneous, but that delivered to organs/tissues is largely inhomogeneous (as well as the dose-rate). In classical radiobiology is it well known that different organs/tissues respond  differently to heterogeneous dose distributions, and we refer organs/tissues as \emph{parallel}, when toxicity is correlated with the mean dose, or \emph{serial}, when toxicity is correlated with the maximum dose in the organ/tissue (most organs/tissues are intermediate between parallel and serial~\cite{mayles2007}.

Because the FLASH effect requires quite large doses to be observed, it may happen that parallel/serial organs respond differently to FLASH-RT when they are irradiated with heterogeneous doses. This may be particularly important if the causes behind the FLASH effect are mostly \emph{local} (\emph{i.e.} affected by the local deposition of dose in a given region, like the physiochemical effects discussed above) rather than \emph{global} or systemic, like the sparing of circulating immune cells. In this situation, it seems natural to argue that the effective maximum dose in an organ/tissue may be more affected by the FLASH effect than the effective mean dose, and therefore the FLASH effect might be more important in serial than in parallel organs/tissues when irradiated by inhomogeneous dose distributions. This has been suggested in the literature, but not fully studied yet.

In this modeling study we qualitatively investigate this hypothesis. In order to do so, we model the FLASH effect as been caused ROD. This was done because the oxygen effect in radiotherapy is well known and there are solid and simple models to account for it through the use of Oxygen Enhacement Ratios (OERs). However, we also employ a phenomenological model linking the local FLASH effect to the dose and the dose rate, and argue that most of the results are qualitatively not dependent on the particular scenario used to model the FLASH effect, and would hold if the mechanism behind the FLASH effect is another as long as it is a local mechanism.

\section{Materials and Methods} \label{section_materials}

\subsection{Radiolytic oxygen depletion and surviving fractions} \label{section_materials_pO2}

We modeled the local effect of FLASH-RT as the effect of the radiolytic oxygen depletion on the surviving fraction of irradiated cells. Following our previous work \cite{gonzalez2024} and Taylor~\textit{et~al.}~\cite{taylor2022}, we modeled the dynamics of oxygen in heterogeneously oxigenated tissues with the following partial differential reaction-diffusion equation:

\begin{equation}
     \frac{\partial p(\mathbf{x},t)}{\partial t} = D_\mathrm{O_2} \Delta p(\mathbf{x},t) - g_\mathrm{max}\frac{p(\mathbf{x},t)}{k+p(\mathbf{x},t)}-\frac{D}{T}G_0\frac{p(\mathbf{x},t)}{k_\mathrm{ROD}+p(\mathbf{x},t)}, \label{eq_oxyFlash}
\end{equation}

In the case of homogeneous oxygenations, we simplified the equation to an ordinary differential equation: 

\begin{equation}
     \frac{{\rm d} p(t)}{{\rm d} t} = -\frac{D}{T}G_0\frac{p(t)}{k_\mathrm{ROD}+p(t)} + \lambda\left( 1-\frac{p(t)}{p(0)}\right) p(t). \label{eq_oxyFlash_2}
\end{equation}

In these equations, $D_\mathrm{O_2}$ is the oxygen diffusion coefficient, $g_\mathrm{max}$ the maximum metabolic consumption rate, $k$ the oxygen pressure for half-maximum consumption rate, $D$ the radiation dose, $T$ the total time of irradiation (the dose rate $R=D/T$), $G_0$ the radiolytic consumption rate, $k_{\rm ROD}$ the oxygen pressure for half-maximum oxygen depletion rate, and $\lambda$ is a recovery rate that controls a logistic recovery term.

These equations can be numerically solved to obtain $p(t)$ (or $p(x,t)$ in the heteregeneous scenario). The value of $p(t)$ can later be used to obtain the effect of a varying oxygen concentration on the surviving fraction. Several methods have been developed for this purpose \cite{petersson2020, taylor2022, song2023, zhu2024}, which were reviewed and extended in \cite{pardo2026}. In this work we used the the LQ-differential method introduced in \cite{pardo2026} to compute the surviving franction in FLASH-RT. This method consists on solving the differential equation:

\begin{equation}
\frac{ {\rm d} \mathit{SF(t)}} {{\rm d}t}  = - \left [ \alpha(t) R + 2 \beta(t) R D(t)\right ] \mathit{SF(t)}
\label{LQ_diff}
\end{equation}
In this equation, $R$ is the dose rate (which we consider constant) and $D(t)$ is the dose delivered up to time $t$ ($D(t)=Rt$). The radiation damage parameters, $\alpha$ and $\beta$, depend on the oxygen partial pressure as \cite{wouters1997}:
\begin{eqnarray}
	&\displaystyle \mathit{\alpha(t)} = \frac{\alpha_{\mathrm{ox}}}{{\mathit{OER}_\alpha}}\frac{{\mathit{OER}_\alpha} \; p(t)  + k_\mathrm{m}}{p(t)  + k_\mathrm{m}}, \label{eq_OERa}\\
	&\displaystyle \mathit{\beta(t)} = \frac{\beta_{\mathrm{ox}}}{{\mathit{OER}_\beta}^2}\frac{({\mathit{OER}_\beta} p(t)  + k_\mathrm{m})^2}{(p(t)  + k_\mathrm{m})^2}.
	\label{eq_OERb}
\end{eqnarray}

Equation (\ref{LQ_diff}) was numerically solved with time-oxygenation curves $p(t)$ obtained from the ROD equations, and the surviving fraction at the end of dose delivery ($t=T$) taken as $\mathit{SF}_\mathrm{F}$).

For CONV-RT ($\mathit{SF}_\mathrm{C}$) ROD is negligible, and the surviving fraction was simply calculated by using the LQ-model:

\begin{equation} \label{eq_LQ}
    \mathit{SF_{\rm{C}}(p,D)} = \exp (-\alpha(p) D-\beta(p) D^2),
\end{equation}
with $\alpha(p)$ and $\beta(p)$ given by equations (\ref{eq_OERa}-\ref{eq_OERb}).

\subsection{LKB model of toxicity and FLASH effective dose}

We used the Lyman-Kutcher-Burman (LKB) model for Normal Tissue Complication Probability (NTCP) \cite{lyman1985, kutcher1989}:

\begin{equation}
\mathit{NTCP} = \frac{1}{2 \pi} \int_{-\infty}^{x} \exp(-t^2/2) {\rm d}t,
\label{eq_LKB}
\end{equation}
where
\begin{equation}
x = \frac{\mathit{gEUD} - \mathit{D_{\rm 50}}} {m \mathit{D_{\rm 50}}}.
\label{eq_LKB2}
\end{equation}

In the above equation, $\mathit{D_{\rm 50}}$ is a parameter corresponding to the dose at which the probability of toxicity is 50\%, $m$ is a parameter that controls the steepness of the toxicity curve. Finally, $\mathit{gEUD}$ is the generalized equivalent uniform dose introduced by Niemierko \cite{niemierko1999}, which for a dose distribution $\{D_i\}$, is given by,

\begin{equation}
\mathit{gEUD} = \left (\sum_i \nu_i D_i^{1/n} \right ) ^n,
\end{equation}
where $\nu_i$ is the weight of voxel $i$ (we will use equi-weights for each voxel, $\nu_i=1/N$ $\forall i$, where $N$ is the number of voxels). The parameter $n$ is the volume efect parameter that controls the response of the organ (\emph{serial} organ in the limit $n \to 0$, \emph{parallel} organ in the limit $n \to 1$).

To compute the NTCP for FLASH-RT we performed a scaling of the delivered dose distribution $\{D_i\}$ to obtain the \emph{effective} FLASH-RT dose, $\{D^{\rm F}_i\}$. Two strategies were followed, one relying on the effect of ROD on the surviving fraction according to the methodology presented in section \ref{section_materials_pO2}, and one relying on a simple phenomenological logistic model. Due to the protective effect of FLASH-RT, $D^{\rm F}_i \leq D_i$.

\begin{itemize}
\item Effective dose: ROD

The \emph{effective} FLASH-RT dose was computed by numerically solving the following equation:
\begin{equation}
D^{\rm F}_i \mid \mathit{SF}_\mathrm{C}(D^{\rm F}_i) = \mathit{SF}_\mathrm{F}(D_i),
\label{eq_DF}
\end{equation}
where $\mathit{SF}_\mathrm{C}$ and $\mathit{SF}_\mathrm{F}$ are the surviving fractions for CONV-RT and FLASH-RT calculated as shown in section \ref{section_materials_pO2}.

\item Effective dose: logistic model

Because the role of ROD on explaining the FLASH effect has been questioned, we also investigated a simple phenomenological equation which models the (local) FLASH effect as a logistic function of the dose and the dose rate. For this we used the Hill's function: 
\begin{equation}
D^{\rm F}_i = D_i \left( 1- A \left( \frac{D_i ^a}{D_i^a + M_D^a} \right) \left( \frac{R_i^b}{R_i^b + M_R ^b}\right) \right).
\label{eq_DF2}
\end{equation}
In this equation, $A$ is a parameter that controls the maximum FLASH effect, $a$ and $b$ and $M_D$ and $M_R$ are parameters that control the slope and dose at 50\% effect for the dose and dose rate dependences, respectively. The inverse of the dose modifying term in the above equation can be viewed as the magnitude so-called FLASH modifying factor (FMF).
\end{itemize}

\subsection{Implementation, parameter values, and dose distribution}

\begin{figure}[t]
	\centering
	\includegraphics[width=12cm]{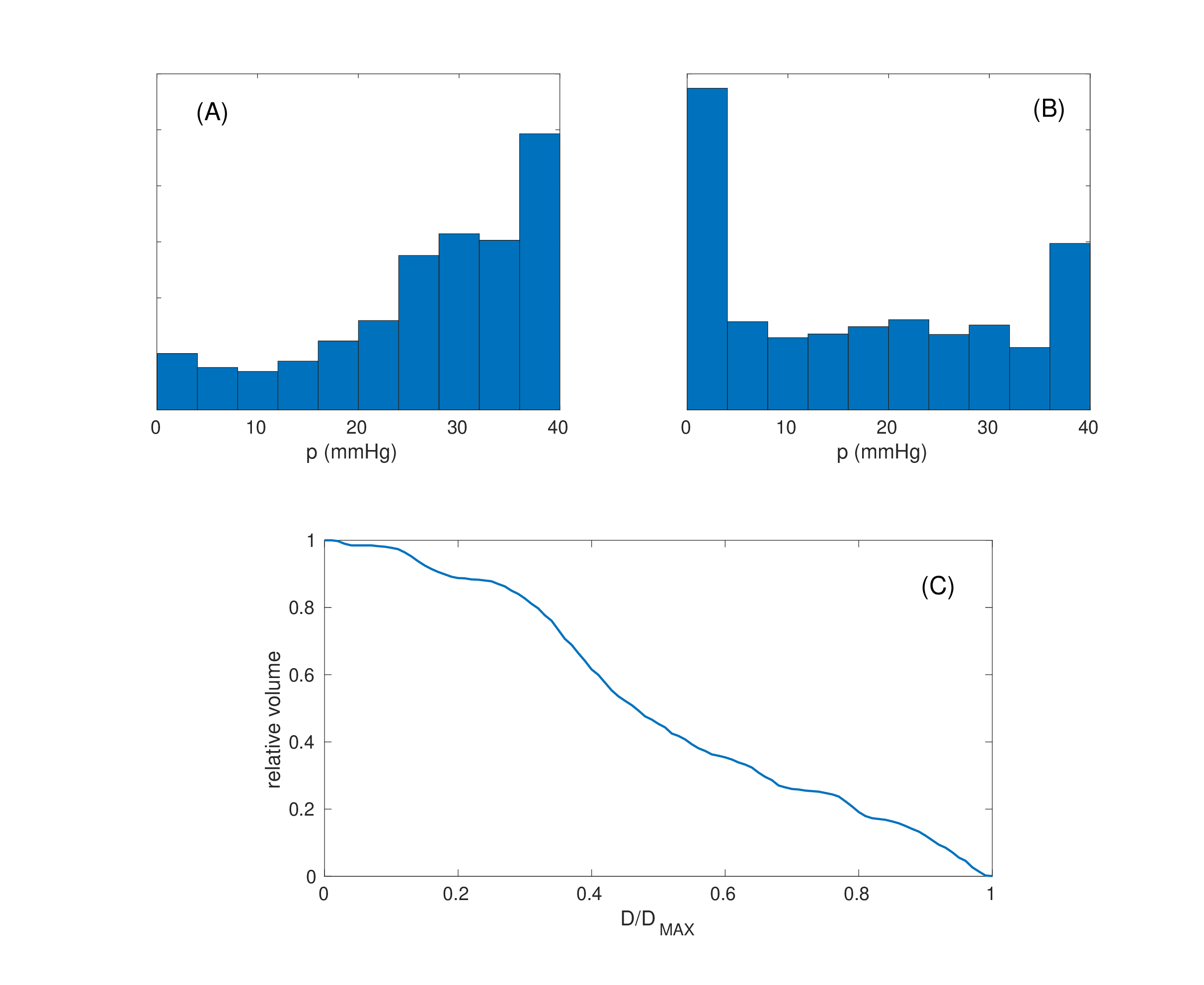}
	\caption{Oxygenation histograms before irradiation for the two heterogeneous oxygen distributions investigated in this work, described as moderately well (A) and poorly (B) oxygenated cells. Dose volume histogram representing the heterogeneous dose distribution used in this work (C).}
	\label{fig1}
\end{figure}

We investigated the effect of FLASH sparing on the NTCP for different values of oxygenation up 30 mmHg. We also investigated the effect for two heterogeneous distributions taken from \cite{gonzalez2024} (more detailed information on the calculation of this distribution can be found in the reference) corresponding to poorly and moderately well oxygenated cells. The baseline oxygenation histograms of these two distributions is shown in Figure \ref{fig1}(A) and (B). Notice that this heterogeneity refers o intra-voxel heterogeneity \cite{petit2009}, and each voxel in the dose distribution $\{D_i\}$ is assumed to have the same oxygen heterogeneity.

In Table \ref{table1} we present a list of the parameters used in this work. ROD dynamics and radiosensitivity parameters are based on those reported in \cite{gonzalez2024}. Regarding the parameters of the LKB model, we used $m$=0.36~\cite{werner2010}, and we investigated several values of $n$ ranging from 1 (\emph{parallel} organ) down to 0.006 (highly \emph{serial} organ). Regarding the parameters used in the phenomenological FLASH-RT effective dose, we have used $M_D$=20 Gy and $M_R$=40 Gy s$^{-1}$ to observe a FLASH effect only at large dose and dose rat values. Also, $a$=2 and $b$=5, to have a steeper curve for the dose rate curve. The FLASH modifying factor, which is the inverse of the effective dose factor calculated in equation (\ref{eq_DF2}), has been experimentally determined to reach values up to 40\%-50\%. Such FMF can be reached by setting $A$=1/3.

\begin{table}[htb]
    \caption{\label{table1} List of parameter values used in this work.}
    \begin{center}
        \begin{tabular}{@{}ll|ll}
                                   \hline
\multicolumn{2}{c|}{\textbf{Radiosensitivity}}  &	\multicolumn{2}{c}{\textbf{NTCP}}    \\
          \hline
$\alpha_{\rm ox}$       & 0.2 Gy$^{-1}$ &  $m$ &  0.36  \\
$\alpha_{\rm ox}/\beta_{\rm ox}$        & 10 Gy   &  $n$ &  (1, 0.7, 0.5, 0.2, 0.1, 0.05, 0.01, 0.006)    \\
$M_\alpha$ & 2.5   & $\mathit{D_{\rm 50}}$ & variable (20 Gy as reference)\\
$M_\beta$  & 2.5         &    $M_D$ & 20 Gy\\
$k_\mathrm{m}$  & 3.28 mmHg        & $M_R$ & 40 Gy s$^{-1}$ \\
$R$          & up to 100 Gy s$^{-1}$ & $a$& 2   \\ \cline{1-2} \cline{1-2}
\multicolumn{2}{c|}{\textbf{ROD dynamics}} & $b$ &5 \\
\cline{1-2}
$D_\mathrm{O_2}$        & 2$\times$10$^{-9}$ m$^2$ s$^{-1}$   & $A$&1/3\\
$g_\mathrm{max}$        & 15 mmHg s$^{-1}$  & & \\
$k_{\rm ROD}$     & 1 mmHg        	   & & \\
$G_0$                   & 0.25 mmHg Gy$^{-1}$	 & & \\
$k$          & 2.5 mmHg  & & \\  
$\lambda$ & 1 s$^{-1}$ & & \\    
           \hline
        \end{tabular}
    \end{center}
\end{table}

%

We used a heterogeneous dose distribution qualitatively representing the dose received by an organ (taken from a previous paper by the authors on treatment planning optimization) to qualitatively investigate the effect of dose heterogeneity on the FLASH effect. The dose volume histogram (DVH) of the distribution is shown in Figure \ref{fig1}(C). The dose was assumed to be delivered in a single fraction (to maximize the FLASH effect). The normalization of the dose distribution (and the definition of $D_{\rm 50}$) was performed as follows:

\begin{enumerate}
  \item For $n$=1 (\emph{parallel} organ) and $D_{\rm 50}$=20 Gy the dose distribution was scaled to yield a desired reference NTCP value (typically 0.2) for CONV-RT.
  \item For the other values of $n$ the same dose distribution of Step 1 was used, but because the value of $n$ is different, the value of $D_{\rm 50}$ was changed to maintain the reference NTCP value (0.2) for CONV-RT.
  \item A dose rate $R$=100 Gy s$^{-1}$ was assigned to the maximum dose in the dose distribution ($D_{\rm MAX}$). The dose rate for the other voxels was scaled as $R_i=100 \frac{D_i}{D_{\rm MAX}}$. 
  \item The same dose distributions (and $n$ and $D_{\rm 50}$ values) were used to calculate the NTCP associated to FLASH-RT by applying the methodology given by equations (\ref{eq_DF}) or (\ref{eq_DF2}).
\end{enumerate}

The change of $D_{\rm 50}$ performed in Step 2 was especifically implemented to investigate the effect of $n$ on the NTCP of FLASH-RT for exactly the same dose distributions (to avoid bias due to employing different dose distributions) and endpoint ($\mathit{NTCP}$=0.2, to avoid bias due to investigating different regions of the NTCP curve).

Equations (\ref{eq_oxyFlash}), (\ref{eq_oxyFlash_2}) and (\ref{LQ_diff}) were solved for a discrete set of dose and dose rate values: $D$ ranging from 0 to 60 Gy in steps of 1 Gy, $R$ ranging from $\simeq$0 (the CONV-RT limit given by equation (\ref{eq_LQ})) to 100 Gy s$^{-1}$ in steps of 5 Gy s$^{-1}$. This dataset was then interpolated and the interpolant used to compute the value of $\mathit{SF}_\mathrm{F}$ for arbitrary values of $D$ and $R$.

The implementation of the models and methodologies was performed in Matlab (Natick, USA). The main function and the data are available in the github repository (see Data availability statement).

\section{Results} \label{section_results}

\begin{figure}[t]
	\centering
	\includegraphics[width=16cm]{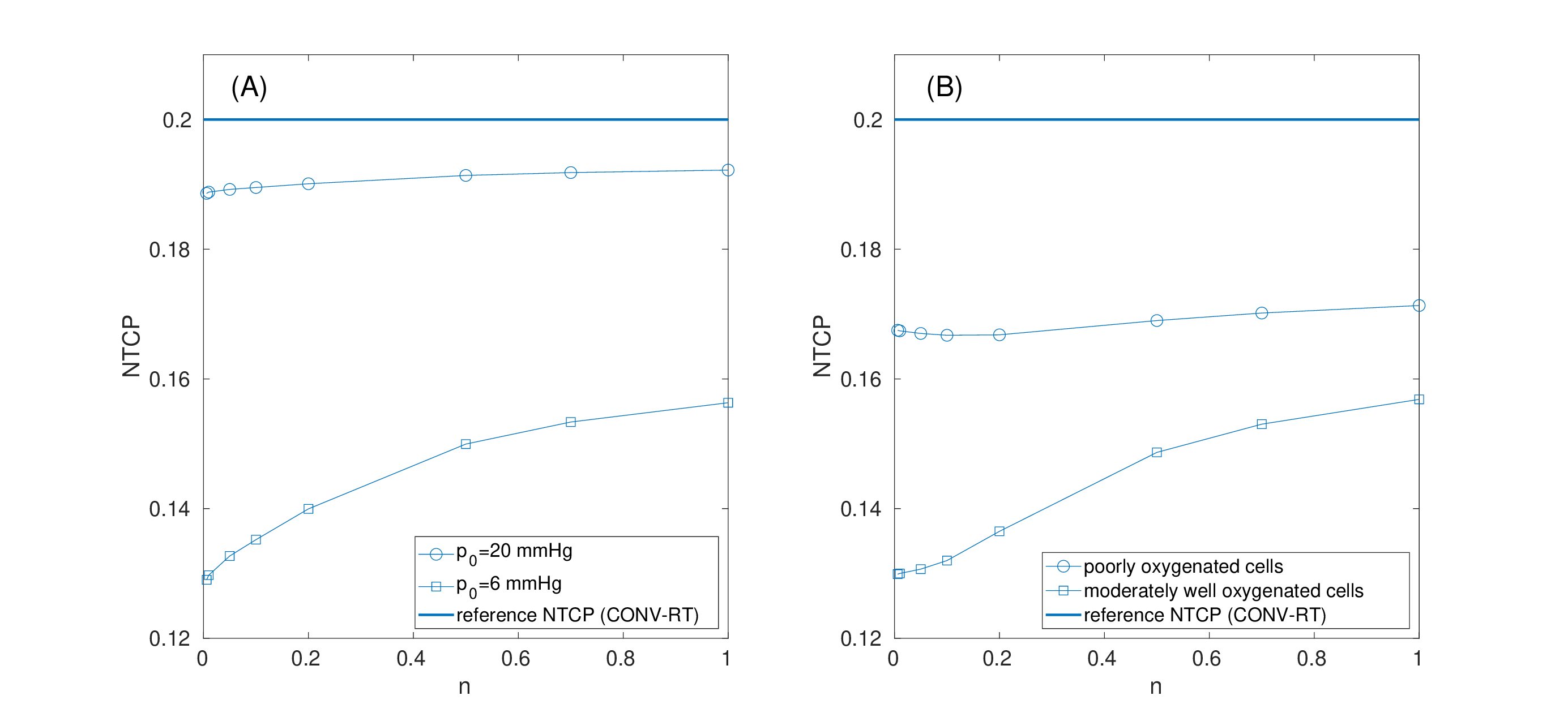}
	\caption{NTCP computed with the LKB model for FLASH-RT assuming that the FLASH modifying factor is due to ROD: (A) results for heterogeneous oxygenations (6 and 20 mmHg); (B) results for heterogeneous oxygenations. The \emph{reference} NTCP (that achieved with CONV-RT) is 0.2.}
	\label{fig2}
\end{figure}

In Figure \ref{fig2}(A) we present the NTCP obtained for FLASH-RT, calculated by assuming that the FLASH dose modifying factor is due to ROD (equation (\ref{eq_DF})), versus the volume parameter $n$ for homegeneous oxygenations (6 and 20 mmHg). The \emph{reference} NTCP, that achieved with CONV-RT, for this calculations is 0.2 (to achieve this NTCP, the maximum dose in the dose distribution shown in Figure \ref{fig1} is 26.79 Gy).

It was observed that FLASH-RT decreases the NTCP, especially for cells with low oxygenations ($\mathit{NTCP}_{\rm FLASH}<$0.16 for 6 mmHg versus $\mathit{NTCP}_{\rm FLASH}$ $>$0.18 for 20 mmHg). This is a very well know result: the radiosensitivity of well oxygenated is not highly affected by ROD because the decrease in oxygenation ($\Delta p \sim$ 5 mmHg) is not enough to reach the oxygenation region where the radiosensitivity of the cells is highly sensitive to oxygenation \cite{taylor2022, gonzalez2024}. More interestingly, the FLASH sparing increased with decreasing $n$, \emph{i.e.} for more serial organs.

The latter trend is confirmed for heterogeneous oxygenations, as shown in Figure \ref{fig2}(B). There are some subtleties in the trends observed here that merit some futher discussion. On the one hand, in this case, the more oxygenated distribution presents a higher FLASH effect. The interplay between dose and FLASH effect in heterogeneously oxygenated organs can be quite complex as discussed in Ref. \cite{taylor2022} (and to some extend also in \cite{gonzalez2024}). The surviving fraction for heterogeneous oxygenations is mostly driven by extremely hypoxic cells ($p \sim0$), and those cells are not affected by ROD because there is almost no oxygen to deplete. On the other hand, we can observe an inversion of the $\mathit{NTCP}_{\rm FLASH}$ vs. $n$ trend at low values of $n$ in the moderately well oxygenated scenario. We attribute this behaviour to the complex interplay between dose and FLASH effect in heterogeneously oxygenations that we previously mentioned, and we refer the reader to cited publications. We did not investigate this (otherwise interesting) effect in detail because such behaviour is intrinsically associated to considering ROD as the driving effect behind the FLASH effect \cite{taylor2022}, but we are only using ROD as a surrogate to investigate a \emph{local} mechanism behind the FLASH effect. We are mostly interested in the general trend of $\mathit{NTCP}_{\rm FLASH}$ vs. $n$, which may traslate to different \emph{local} mechanisms, rather than in particular trends associated to ROD, which may not translate.

\begin{figure}[t]
	\centering
	\includegraphics[width=16cm]{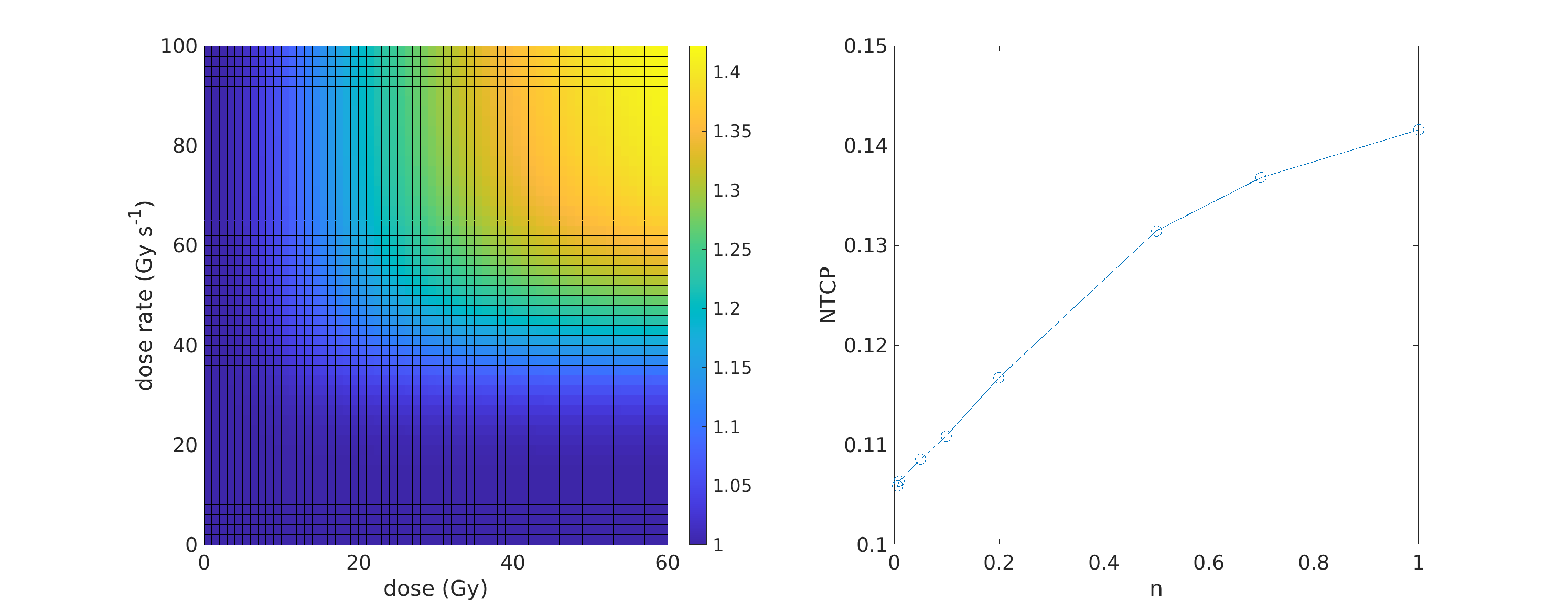}
	\caption{Left: FLASH modifying factor (FMF), computed as the inverse of the effective dose factor reported in equation (\ref{eq_DF2}) with the parameters reported in Table \ref{table1}, versus dose and dose rate. Rigth: NTCP computed with the LKB model for FLASH-RT assuming the FMF shown in the left panel versus the volume parameter of the organ/tissue. The \emph{reference} NTCP (that achieved with CONV-RT) is 0.2.}
	\label{fig3}
\end{figure}

In Figure \ref{fig3} we show the results obtained when employing a simple phenomenological model to account for the FLASH dose modifying factor. In the left panel we shown the modeled dependence of FMF on the dose and dose rate (defined as the inverse of the dose modifying factor in equation (\ref{eq_DF2})), which computed with the parameters reported in Table \ref{table1} reaches up to 1.5 for very large doses and dose rates. In the left panel we show the computed values of $\mathit{NTCP}_{\rm FLASH}$ versus the volume parameter $n$. Again, there was significant sparing associated to the irradiation with FLASH-RT ($\mathit{NTCP}_{\rm FLASH}<$0.15, and also a clear trend for which serial organs are more sensitive to the FLASH-RT sparing that paralell organs ($\mathit{NTCP}_{\rm FLASH}\simeq$0.14 for $n$=1 versus $\mathit{NTCP}_{\rm FLASH}\simeq$0.11 for $n$=0.1).

\section{Discussion} \label{section_discussion}

That FLASH radiotherapy (FLASH-RT) can spare normal tissue compared to conventional radiotherapy (CONV-RT) is well established \textit{in vivo}. But he mechanisms behind this sparing are still not entirely clear, with different studies pointing to radiolytic oxygen depletion (ROD), variations in reactive species production, recombination of free radicals, or lipid metabolism or radiation-induced immune effects. 

The so-called FLASH effect is dose and dose rate dependent, necesitating of large dose rates ($\sim$ 40 Gy $^{-1}$) and large doses ($\sim$ 10 Gy) to be observed. Experimental studies typically involve the irradiation of an organ with quite homogeneous doses. However, in a clinical scenario, the dose (and the dose rate) received by irradiated organs/tumors is quite inhomogeneous, with subvolumes receiving doses as large as that in the tumor (for organs abutting the tumor), and other subvolumes receiving low doses. Because the FLASH effect requires quite large doses and dose rates to be observed, parallel/serial organs may respond differently to FLASH-RT when they are irradiated with heterogeneous doses. This may be particularly important if the causes behind the FLASH effect are mostly \emph{local} (\emph{i.e.} affected by the local deposition of dose in a given region, like the physiochemical effects discussed above) rather than \emph{global} or systemic, like the sparing of circulating immune cells. In this situation, it seems natural to argue that the effective maximum dose in an organ/tissue may be more affected by the FLASH effect than the effective mean dose, and therefore the FLASH effect might be more important in serial than in parallel organs/tissues when irradiated by inhomogeneous dose distributions.

In this modeling study, we qualitatively investigated this hypothesis. To do so, we modeled the \emph{local} FLASH effect in two ways: i) as caused by radiolytic oxygen depletion, and; ii) with a purely phenomenological model that links the FLASH effect to dose and dose rate. The former model was used because the oxygen effect in radiotherapy is well understood and there are solid and simple models to account for it. However, because it has been argued that ROD cannot be the only cause of the FLASH effect, and because the role of oxygenation can be tricky at times \cite{taylor2022, gonzalez2024}, we also investigated the phenomenological model.

This modeling study seems to support that the biological benefit of FLASH radiotherapy is not only a function of dose and dose rate but also of the spatial architecture of the irradiated organ/tissue. Our modeling indicates that serial organs, whose toxicity is driven by the maximum dose they receive, are likely to benefit more from the FLASH effect than parallel organs, whose toxicity is driven by the mean dose they receive. While some of the results presented in this paper may only hold if ROD is the underlying mechanism of the FLASH effect, the qualitative results of this study should hold true for any underlying mechanism that acts locally.

While this study provides a theoretical basis for the impact of organ architecture on FLASH sparing, it has several limitations that need to be addressed, among them, the particular models employed in the calculations and the particular set of parameters used. Some of these limitations have been addressed above, in particular the use of ROD. Another important limitation is the dose rate distribution used in this work: our model assumes a linear scaling between local dose and dose rate ($R_i \propto D_i$) and a continuous delivery of the dose (no complex time structure). This may be adequate to describe a single wide-field irradiation typical of electron FLASH, or even transmission proton therapy. For different deliveries, the dose rate distribution can be significantly more complex, and there is an interplay between dose, dose rate structure and the FLASH effect \cite{karsch2022, liu2025} which might affect the conclusions of this work. While we expect the qualitative conclussions of this work to hold even for those complex deliveries and dose rate spatial and temporal distributions, they may create a more complex interaction between the FLASH effect, dose and dose rate. This merits further study and will be analyzed in future work.


\section*{Data availability statement}

Data and code are available in the github repository: \url{https://github.com/juancho-pm/FLASH_NTCP}.

\section*{Acknowledgments}

This work has received funding from Xunta de Galicia-GAIN (grant IN607D2022/02) and Ministerio de Ciencia e Innovación-AEI and FEDER, UE (grants PLEC2022-009476 and PID2021-128984OB-I00).


\bibliographystyle{unsrt}
%

\end{document}